\documentclass[a4paper,11pt]{article}

\usepackage[english]{babel}
\usepackage[utf8]{inputenc} 

\usepackage{amsmath,amssymb}
\usepackage{graphicx}
\usepackage[colorlinks=true, allcolors=black]{hyperref}
\usepackage{geometry}
\usepackage{natbib}
\usepackage{graphics}
\usepackage{setspace}

\begin{document}

\title{The Paradox of Doom: Acknowledging Extinction Risk Reduces the Incentive to Prevent It\thanks{We grateful to Pawe{\l} Struski (GRAPE, Warsaw, Poland) for his participation in this project as research assistant. We acknowledge the research funding from the National Science Center, Poland, through the grant ``Will Artificial General Intelligence Bring Extinction or Cornucopia? Modeling the Economy at Technological Singularity'' (OPUS 26 No. 2023/51/B/HS4/00096).}}


\author{Jakub Growiec\thanks{SGH Warsaw School of Economics, Poland and CEPR Research and Policy Network on AI.}  \and Klaus Prettner\thanks{WU Vienna, Austria.}}

\maketitle


\vspace{30mm}

\begin{abstract} We investigate the salience of extinction risk as a source of impatience. Our framework distinguishes between human extinction risk and individual mortality risk while allowing for various degrees of intergenerational altruism. Additionally, we consider the evolutionarily motivated ``selfish gene'' perspective. We find that the risk of human extinction is an indispensable component of the discount rate, whereas individual mortality risk can be hedged against---partially or fully, depending on the setup---through human reproduction. Overall, we show that in the face of extinction risk, people become more impatient rather than more farsighted. Thus, the greater the threat of extinction, the less incentive there is to invest in avoiding it. Our framework can help explain why humanity consistently underinvests in mitigation of catastrophic risks, ranging from climate change mitigation, via pandemic prevention, to addressing the emerging risks of transformative artificial intelligence.


\end{abstract}

\newpage

\section{Introduction}


Each of us is going to die, but we had better not all die on the same day. For it is the asynchronicity of human deaths that sustains our species and keeps the human civilization going. Across a sequence of overlapping generations of mortal humans, the human species as a whole may appear immortal. Unfortunately, more than 99 percent of all species that ever lived on Earth are estimated to be extinct now,\footnote{\url{https://en.wikipedia.org/wiki/Global_biodiversity} [accessed: 29.08.2025].} and one day humans may face  the same fate. 


How should we think about the perspective of human extinction? On the one hand, human extinction should intuitively imply a greater loss than the combined loss of all lives, because it would also entail the loss of human history and heritage, as well as our potentially glorious future. But, on the other hand, perhaps the value of the past lies solely in providing opportunities for the humans alive today? And perhaps the value of the future can be considered only when a given future actually came to pass, which is \emph{a priori} uncertain and, in expectation, already accounted for in present utility? So, from the perspective of each individual human, what is the actual difference between individual death and the disappearance of humanity as a whole?

The purpose of this paper is to thoroughly inspect the theoretical link between the prospects of human extinction and discounting. Mortality risk is known to be one of the key reasons why we discount the future: if survival is uncertain, one prefers to consume earlier rather than later---to avoid dying in the meantime and not being able to consume at all \citep{Fisher1930,RobsonSamuelson2009}. The same logic works for the risk of human extinction: it is better to consume early, before the extinction risk materializes. But it is unclear whether individual mortality or aggregate extinction risk exert a stronger impact on people's impatience. On the one hand, individual mortality is tangible, whereas extinction risk may seem distant, speculative, and therefore easier to ignore. On the other hand, unless one's preferences are purely selfish, mortality risk---as opposed to extinction risk---can be at least partially hedged against through reproduction. The phrase \emph{non omnis moriar,} ``not all of me will die'', expressing the idea that one's legacy endures through one's descendants, illustrates exactly this sentiment. 

In this study we construct a tractable framework in which people's impatience stems from both idiosyncratic (individual mortality) risk and aggregate (human extinction) risk. We carry out the risk calculation from the perspective of a single, selfish, and finite-lived individual; a dynasty; and a social planner who maximizes cumulative social welfare. When aggregating individual utilities, we distinguish between aggregate vs. average utilitarianism \citep{Parfit1986}. Additionally, we also consider the evolutionary ``selfish gene'' perspective \citep{Dawkins1976}.  Doing so allows us to illustrate the difference between mortality risk and extinction risk and analyze the consequences that a changing risk of extinction may have on our behavior. 




We find that across all considered scenarios, the risk of human extinction is always a source of impatience, whereas individual mortality risk can be (at least partially) alleviated through human reproduction. If there is a positive degree of intergenerational altruism---which also arises in the ``selfish gene'' perspective---then postponing consumption is not so strongly affected by individual mortality risk, because even if one does not survive long enough to consume individually, one still obtains some utility from the fact that one's offspring will consume.

Our study links to a literature on the economics of catastrophes. This topic has been of particular interest among environmental economists dealing with possible consequences of climate change \citep{Weitzman2012,MartinPindyck2015,ChichilniskyEtal2020}. Scholars have, in particular, developed theories allowing to include extinction risk in the endogenous aggregate discount rate \citep{Chichilnisky2000}. 
Our analysis is also closely linked to the literature on evolutionary foundations of intertemporal preferences, specifically in the presence of both idiosyncratic and aggregate (though not extinction-level) uncertainty \citep{Robson2001,RobsonSamuelson2009,RobsonSamuelson2010}. Against this background, we observe that the theoretical distinction between individual mortality risk and extinction risk is still insufficiently explored. 

The timeliness of this research stems from various developments in areas that are associated with risks of human extinction, for example, the recent rapid progress towards superhuman transformative artificial intelligence (AI) algorithms, which have been argued to pose imminent extinction risks to humanity, either through accident, misuse by bad actors, or through AI takeover by a misaligned superhuman AI \citep{Bostrom2014,Jones2023,GrowiecPrettner2025,YudkowskySoares2025}; a changing climate beyond crucial tipping points that could generate positive feedback effects between past and future warming (such as melting polar ice caps that reduce the cooling albedo effect) and between warming and further greenhouse gas emissions (such as the methane released from melting permafrost); the threat of a war between nuclear weapon states; or the occurrence of a global pandemic deadlier than COVID-19. These new risks add to other sources of ``background'' extinction risk such as an asteroid impact, a supervolcano eruption, etc. \citep[cf.][]{Ord2020}. 

This rapidly emerging new context calls for answering the following questions: 
\begin{itemize}
\item If, based on the new information, people will update their beliefs on the underlying extinction risk upwards, will they become more impatient? 
\item If so, will this increase in impatience be smaller or greater than the increase in impatience observed among individuals who update their beliefs about their own expected lifespan, for example, when they are diagnosed with a terminal illness?
\item If rising extinction risks lead to higher discounting, what does this imply for investment in the prevention of extinction in the first place?
\end{itemize}

Overall, we find that extinction risk is an important contributor to impatience that cannot be hedged against from an individual perspective. Thus, impatience will inevitably rise when the  risk of extinction is acknowledged---or when people's beliefs on the underlying extinction risk are corrected upwards---which leads to the paradoxical situation that investments in mitigating catastrophes become less attractive with a rise in impatience through extinction risk. This, in turn, makes extinction even more likely. Our results provide an additional explanation  why it is so hard to promote investments in mitigating human extinction risk, separate from the observation that this risk is often dismissed as speculative and empirically unjustified based solely on the grounds that thus far humanity has not died off yet. 

The structure of the paper is as follows. We review the literature and important concepts in Section \ref{sec:lit}. Section \ref{sec:model} contains the framework that we use to disentangle individual mortality risk from extinction risk.  In Section \ref{sec:discussion}, we discuss the implications of our findings. In Section \ref{sec:conclusions}, we conclude.

%

\section{Literature Review}
\label{sec:lit}

\subsection{Motives for Discounting}

It has been thoroughly documented that people discount the future \citep{FrederickEtal2002,CohenEtal2020}. Famous ``marshmallow tests'' have demonstrated that already four-year old children have consistent intertemporal preferences, with varying willingness to trade off one unit of a consumption good (such as a marshmallow) now for two units after 15 minutes \citep{MischelEtal1989}. According to a meta-analysis of micro-level experimental evidence carried out by \cite{MatousekEtal2022}, adult people's mean annual discount rate is about 0.33, and we tend to be less patient when health is at stake compared to money. However, the same study also emphasizes massive uncertainty surrounding the estimates of individual discount rates and their dependence on context. For example, people appear to be more patient when surrounded by natural as compared to urban landscapes \citep{VanDerWalEtal2013}.\footnote{There is also a voluminous literature documenting that human preferences are time inconsistent and exhibit present bias. Discounting is probably better described by a hyperbolic than an exponential curve \citep{ImaiEtal2021}. Addressing this issue is beyond the scope of our paper, though.}

By contrast, discount rates used in the calibration of macroeconomic models, or those estimated based on aggregative data, are much lower and typically vary around 0.05 per annum \citep[see e.g.][]{CooleyPrescott1995,GommeRupert2007}. This discrepancy between experimental evidence on individual-level impatience and aggregate discount rates, which ought to be consistent with economy-wide interest rates and capital accumulation rates, suggests that intertemporal preferences may not straightforwardly aggregate in the human population.

Our willingness to discount the future is likely a consequence of our evolutionary history. After all, impatience is also observed across a wide range of animal species \citep{VanderveldtEtal2016,Hayden2016}. In an unpredictable natural environment, where one cannot be certain about the future states of the world, moderate amounts of greed and myopia deliver clear evolutionary advantages. And uncertainty pertains also to each individual's lifespan: given that survival is uncertain, one prefers to consume earlier rather than later---to avoid dying in the meantime and not being able to consume at all \citep{Fisher1930,RobsonSamuelson2009}.

However, people's preferences are usually not entirely selfish; instead we often exhibit sub\-stantial amounts of intergenerational altruism. We are motivated to have children, leave bequests to them, and draw utility from their consumption beside our own. Through this evolutionary lens, by reproduction, our ``selfish genes'' \citep{Dawkins1976} are carried and supported across generations. But while we care for our children, their consumption matters for our utility not quite as much as our own: ``offspring (...) are only half relatives under sexual reproduction (...) [which constitutes] a source of impatience'' \citep{Robson2001}.

\subsection{Extinction Risk and Discounting}

At this point, we would like to ask: how does the perspective of human extinction change this picture? Would acknowledging the risk of extinction drive our discounting behavior in the same way as individual mortality risk does, or would it work differently? The literature on existential\footnote{Existential and extinction risks are often perceived as synonyms, but the category of existential risks is actually slightly wider. According to \cite{Bostrom2002}, existential risk is the risk that ``(...) an adverse outcome would either annihilate Earth-originating intelligent life or permanently and drastically curtail its potential''. This includes also, e.g., scenarios where misaligned superhuman AI does not make humans literally extinct, but nevertheless takes over and permanently disempowers them. In the remainder of this paper, we will be referring to extinction risks exclusively.} and other catastrophic risks  \citep{Chichilnisky2000,Weitzman2012,MartinPindyck2015,ChichilniskyEtal2020} is scarce and cannot offer a comprehensive answer to this question. We see at least two separate mechanisms here.

First, there is a possibility that the risk of human extinction---despite being very real---will be entirely dismissed on the basis of lack of historical precedents. Indeed, this seems to be the usual stance in research practice and policy debates: risks that have never materialized in the past, and, thus, cannot be empirically studied, are easily dismissed as unjustified and speculative \citep[as discussed by][in the context of risks from superhuman AI]{GrowiecPrettner2025}. Worryingly, this logic is applicable regardless of the severity of the true underlying extinction risk, and despite that, by definition, if that risk materializes, there will be no more humans to observe it.

The second mechanism arises once the extinction risk is acknowledged. Internalizing an additional source of mortality risk means that, other things equal, one's expected lifespan goes down, and hence one's impatience should increase. 
Specifically, from the point of view of a utility-maximizing individual with purely selfish preferences, there should be no difference between idiosyncratic vs. aggregate risk (own death vs. human extinction) because both of them affect the individual's lifespan in exactly the same way. However, with intergenerational altruism---or in the evolutionary ``selfish gene'' perspective---extinction risk could induce \emph{more} impatience than individual deaths because idiosyncratic mortality risks can be hedged through reproduction, whereas aggregate risks cannot. As \cite{Robson2001} put it for the case of correlated but non-extinction risks: ``a gamble that is idiosyncratic is strictly preferred to one that is aggregate, even if the two distributions are identical''. This logic works for extinction risks as well, which is particularly easy to see if intergenerational altruism emerges from the optimization problem of the ``selfish gene'': non-hedgeable extinction risks are fully discounted because no genetic lineage can survive and multiply if the species is extinct.


\section{The Model}
\label{sec:model}

Human mortality is one of the key justifications for discounting. Specifically, \cite{Blanchard1985}'s elegant (even if not fully realistic) ``perpetual youth'' model shows that even in the absence of other motives for discounting of the future, mortality alone can give rise to exponential discounting on behalf of the optimizing individual, dynasty, or social planner. The same functional form has also been derived from a model with extinction risk \citep{Chichilnisky2000}. We shall now study the emergence of discounting in a tractable model extending \cite{Blanchard1985}, in which there is both individual mortality and extinction risk. For brevity but without loss of generality, we abstract from any other sources of discounting behavior. 

\subsection{Individual Expected Utility}

Consider a population of individuals who live for a finite but uncertain lifetime with time $t$ being treated as discrete: $t=0,1,2,...$. Let us assume we face a constant extinction risk $M\geq 0$ at each point in time, and, as long as humanity survives, a constant individual death risk $m\geq 0$. The constant death hazard is independent of an  individual's age (hence the term ``perpetual youth''). 
 
Let $D$ be a discrete random variable capturing the duration of one's life. An individual born at time $t=0$ will die at time $t\geq 0$ with probability:
\begin{eqnarray}
P(D=0)&=&M+(1-M)m=M+m-Mm, \nonumber \\
P(D=1)&=&(1-m)(1-M)(M+m-Mm), \nonumber \\
P(D=2)&=&(1-m)^2(1-M)^2(M+m-Mm), \\
 &...& \nonumber \\
 P(D=t)&=&(1-m)^t (1-M)^t (M+m-Mm). \nonumber 
\end{eqnarray}


To inspect the importance of individual mortality and extinction risk for discounting under \emph{purely selfish preferences}, we calculate the expected value of lifetime utility of consumption (when there is no time discounting on top of mortality and extinction risks) of a representative individual born at $t=$0, denoted as $U_0$:
\begin{eqnarray}
EU_0 &=& \sum_{t=0}^\infty P(D=t) (u(c_0)+...+u(c_t)) = \sum_{t=0}^\infty (1-m)^t (1-M)^t u(c_t),
\end{eqnarray}
where $c_t$ is consumption at time $t$ and $u(\cdot)$ is the utility function. The discount rate is the sum of the individual death rate $m$ and the aggregate extinction risk $M$: selfish individuals are impatient because they are aware of their mortality, whether idiosyncratic or aggregate. They know that postponing consumption runs the risk of not surviving to the time of consumption, which reduces total utility. But they do not differentiate between individual and aggregate death risk.

\subsection{Expected Utility of a Dynasty}

We will now introduce \emph{intergenerational altruism} to the individuals' preferences, so that instead of evaluating their own individual utility, they calculate the utility of their entire dynasty (or local community, nation, etc.). To calculate the expected utility of a dynasty, we need two additional elements. First, we must make an assumption on the birth rate (and, thus, the growth rate of the dynasty). Assuming a constant birth rate $b>0$, and normalizing $N_0=1$, the dynasty's size grows or declines geometrically from $t=0$ until time $T$ when the extinction risk materializes:
\begin{equation}
N_t=
\begin{cases}
(1+b)^t (1-m)^t = (1+n)^t, &t=0,...,T, \\
0, &t>T.
\end{cases}
\end{equation}
This calculation implicitly assumes smooth temporal evolution of the dynasty's size, thereby excluding the scenario where the dynasty randomly dies off before $T$. This is an innocuous assumption as long as the dynasty's size is sufficiently large.

When births fully hedge the dynasty against the risk of death, the only remaining risk the dynasty faces is the risk of human extinction. As above, we assume that this risk exhibits a constant hazard rate across time, so that the probability of human extinction at any time $T$ is given by 
\begin{equation}
P_X(T)=(1-M)^T M.
\end{equation}

We can then calculate the expected value of total utility of consumption of a representative dynasty $V$  spawned at $t=0$, denoted as $V_0$ (the Benthamite utility function) as
\begin{equation}
EV_0 = \sum_{t=0}^\infty P_X(t) (u(c_0)+(1+n)u(c_1)+...+(1+n)^t u(c_t)) = \sum_{t=0}^\infty  (1-M)^t (1+n)^t u(c_t).
\end{equation}
$EV_0$ is finite as long as $(1-M)(1+n)<1$, which means that unbounded growth in the dynasty's size does not dominate the extinction risk in expectation.\footnote{Please note that, while this may be violated for very low extinction risk $M$, in a general setting with other sources for discounting than individual death and extinction risk it would likely be fulfilled.}

It follows that the dynasty's discount rate is the sum of the individual death rate $m$ and the aggregate extinction risk $M$, minus the birth rate $b$. Specifically, if the dynasty size is constant over time, so that $n=0$ (i.e., $(1+b)(1-m)=1$), then the discount rate is equal to the extinction hazard rate $M$---the only source of uncertainty in a setup where each dying generation is guaranteed to be replaced by a new one.

\subsection{Aggregate vs. Average Utilitarianism}

So far, when evaluating the expected utility of a dynasty we used the Benthamite utility function, i.e., we weighted all generations equally. This is not the only option, however: in reality, a systematic bias is observed according to which utility of people alive today is given more weight than the utility of future generations \citep[][]{Feng2018}. It may be partly the case because under extinction risk, some of the future generations may never come to life; however, it may also be partly the case because individuals' preferences are closer to average utilitarianism (Millian utility) than aggregate utilitarianism (Benthamite utility), cf. \cite{Parfit1986}.


To incorporate this possibility in the calculations of expected utility of a dynasty, we add a parameter $\theta\in[0,1]$, spanning the entire space of intermediate options between average utilitarianism ($\theta=0$), when we only care about average utility in each period of time $t$, and aggregate utilitarianism ($\theta=1$).

We again calculate the expected value of utility of consumption of a representative dynasty $V$  spawned at $t=0$, denoted as $V_0(\theta)$:
\begin{equation}
EV_0 (\theta) = \sum_{t=0}^\infty P_X(t) (u(c_0)+(1+n)^\theta u(c_1)+...+(1+n)^{\theta t} u(c_t)) = \sum_{t=0}^\infty  (1-M)^t (1+n)^{\theta t} u(c_t).
\end{equation}
$EV_0(\theta)$ is finite as long as $(1-M)(1+n)^\theta<1$, which means that unbounded growth in the dynasty's size does not dominate the extinction risk in expectation.

Now the dynasty's discount rate is the difference between the aggregate extinction risk $M$, and the population growth rate (the birth rate $b$ minus the individual mortality rate $m$), taken to the power $\theta$. If the dynasty size is constant over time, the discount rate is again equal to the extinction hazard rate $M$. In turn, if the dynasty is growing over time ($n>0$), then with $\theta<1$, the discount rate is higher than in the case of aggregate utilitarianism ($\theta=1$); conversely, if the dynasty is declining in size, the discount rate is lower than in the case of aggregate utilitarianism.

\subsection{The ``Selfish Gene'' Perspective}

We will now calculate the expected utility of a dynasty using the ``selfish gene'' perspective \citep{Dawkins1976}, which imposes an additional weighting of consecutive generations' utility. From the point of view of an individual genome, utility of next generations should not be worth as much as utility of the present generation, because, under sexual reproduction, genetic code is recombined---the next generation carries only half of the original genome. Moreover, there are minor additional losses due to mutation. 
This puts a wedge between the birth rate $b$ and the mortality rate $m$: when a person dies, so does their genome; when a new person is born, only (slightly less than) half of the genome is reproduced.

Consequently, assuming a constant birth rate $b>0$, and normalizing the size of the genetic lineage to $\Gamma_0=1$, we posit that the ``selfish gene'' spreads geometrically from $t=0$ until time $T$ when the extinction risk materializes:
\begin{equation}
\Gamma_t=
\begin{cases}
(1+b)^{\alpha t} (1-m)^t, &t=0,...,T, \\
0, &t>T.
\end{cases}
\end{equation}
The exogenous parameter $\alpha\in(0,1)$ captures the loss of genetic code due to recombination and mutation, and depends on the average age of human reproduction. By assuming that the gene is sufficiently widespread, we exclude the scenario in which the given genetic lineage randomly dies off before $T$, which is analogous to our assumption with respect to the population size made above. 

We can now calculate the expected value of total utility of consumption of a representative genetic lineage $G$  spawned at $t=0$, denoted as $G_0$:
\begin{eqnarray}
EG_0 &=& \sum_{t=0}^\infty P_X(t) (u(c_0)+(1+b)^\alpha (1-m) u(c_1)+...+(1+b)^{\alpha t} (1-m)^t u(c_t)) = \nonumber \\ 
&=& \sum_{t=0}^\infty  (1-M)^t (1+b)^{\alpha t} (1-m)^t u(c_t).
\end{eqnarray}
$EG_0$ is finite as long as $(1-M)(1+b)^\alpha (1-m)<1$, which means that unbounded growth in the genetic lineage does not dominate the extinction risk in expectation.

We find that the ``selfish gene'' discounts the future unambiguously more aggressively than the Benthamite utilitarian: the discount factor is now additionally multiplied by $(1+b)^{\alpha-1}<1$.

\subsection{Expected Utility of Humanity}

\subsubsection{Individual Expected Utility when the Extinction Date Is Known}

We shall now proceed to calculating aggregate welfare of the entire population. To this end, we will first calculate individual expected utility with a \emph{known and given} extinction date $T$.

In line with the ``perpetual youth'' assumption, we posit that each individual born at time $t=0$ will die at time $t = 0, 1, ..., T$ with probability:
\begin{eqnarray}
P(D=0)&=&m, \nonumber \\
 P(D=1)&=&(1-m)m, \nonumber \\
 P(D=2)&=&(1-m)^2 m, \\
 &...& \nonumber \\
 P(D=T)&=&(1-m)^T m + (1-m)^{T+1} = (1-m)^T, \nonumber 
\end{eqnarray}
where the last term $(1-m)^{T+1}$ pertains to the individuals who did not die from natural reasons but perish at $T$ when the extinction risk materializes.

Hence,
\begin{eqnarray}
EU_0 \Big\rvert_{T -\textrm{given}}&=& \sum_{t=0}^T P(D=t) (u(c_0)+...+u(c_t)) = \sum_{t=0}^T (1-m)^t u(c_t).
\end{eqnarray}
This implies that when the extinction date is known, the discount rate reflects only the individual death rate $m$---i.e., only the uncertainty in one's lifespan, not its expected length. Indeed, the discount rate is exactly the same regardless of the extinction date $T$, which means it disregards the prospects of future extinction entirely.

\subsubsection{Aggregate Social Welfare}

The calculation of the expected utility of the entire human population---aggregate social welfare---proceeds as follows. First, we assume that the population size grows or declines geometrically from $t=0$ until $t=T$ when the extinction risk materializes:
\begin{equation}
N_t=
\begin{cases}
N_0 (1+b)^t (1-m)^t = N_0 (1+n)^t, &t=0,...,T, \\
0, &t>T.
\end{cases}
\end{equation}

Likewise, with a constant extinction hazard across time, the probability of human extinction at any given time $T$ amounts to 
\begin{equation}
P_X(T)=(1-M)^T M.
\end{equation}

The sum of utilities of all individuals born between time 0 and $T$ is
\begin{eqnarray}
W(0,T) &=& \sum_{t=0}^T N_t EU_t \Big\rvert_{T} = \sum_{t=0}^T N_0 (1+n)^t \sum_{\tau=t}^T (1-m)^{\tau-t} u(c_\tau) = \\
&=& N_0 \sum_{t=0}^T u(c_t) (1+n)^t \frac{1-\left( \frac{1-m}{1+n}\right)^{t+1}}{1- \frac{1-m}{1+n}} 
=  N_0 \frac{1+b}{b} \sum_{t=0}^T u(c_t) (1+n)^t \left( 1-\left( \frac{1}{1+b}\right)^{t+1} \right), \nonumber
 \end{eqnarray}
where the last equality follows from the fact that $1+n=(1+b)(1-m)$.

Therefore the expected value of humanity's utility of consumption from period 0 to $+\infty$, i.e., of the social welfare function $W$, is
\begin{eqnarray}
EW &=& \sum_{T=0}^\infty P_X(T) W(0,T) = M N_0 \frac{1+b}{b} \sum_{T=0}^\infty  \sum_{t=0}^T (1-M)^T  u(c_t) (1+n)^t \left( 1-\left( \frac{1}{1+b}\right)^{t+1} \right) = \nonumber \\
&=& M N_0 \frac{1+b}{b} \sum_{t=0}^\infty  u(c_t) (1+n)^t \left( 1-\left( \frac{1}{1+b}\right)^{t+1} \right) \sum_{T=t}^\infty (1-M)^T \\ 
&=& N_0 \frac{1+b}{b} \sum_{t=0}^\infty  u(c_t) (1-M)^t (1-m)^t (1+b)^t  \left(1-\left(\frac{1}{1+b}\right)^{t+1} \right), \nonumber
\end{eqnarray}
which simplifies to 
\begin{eqnarray}
EW &=& \frac{N_0}{ m } \sum_{t=0}^\infty  u(c_t) (1-M)^t \left( 1-(1-m)^{t+1} \right)
\end{eqnarray}
if the population size at $t=0,...,T$ is constant, i.e., $n=0$. Again, the expected value $EW$ is finite as long as $(1-M)(1+n)<1$, which means that unbounded population growth does not dominate the extinction risk in expectation.

We find that the discount rate is no longer constant because the finite-time human extinction violates the ``perpetual youth'' assumption. However, for large $t$ it converges to a constant. Specifically, the long-run discount factor is equal to $(1-M)(1-m)(1+b)<1$. This means---exactly like in the case of the dynasty---that the discount rate is the sum of the individual death rate $m$ and the extinction risk $M$, minus the birth rate $b$.

%

%
%

\section{Discussion}
\label{sec:discussion}

The results of our analysis are summarized in Table \ref{tab:results}. We find that in the case of purely selfish preferences, the individual mortality rate $m$ and the risk of human extinction $M$ are symmetric sources of people's impatience. With intergenerational altruism, however, individual mortality risk is hedged against through human reproduction. 

Under the assumption of a constant population size, the discount factor of an optimizing head of dynasty, or benevolent social planner, reflects solely the extinction risk $M$. Under intergenerational altruism, with a constant population and without other sources of impatience, postponing consumption is unaffected by individual mortality risk because even if one does not survive long enough to consume individually, one still obtains exactly the same utility from the fact that one's offspring will consume.

By contrast, under the evolutionarily motivated ``selfish gene'' perspective, the individual mortality rate matters for the discount factor even if the population size is constant (albeit less than in the purely selfish case). This is because from the perspective of a genome, births and deaths are not symmetric: when a person dies, so does their genome; when a new person is born, only (slightly less than) half of the genome is reproduced. In result, one-to-one reproduction does not fully hedge the genome against individual mortality risk.

We also emphasize that it is not the actual hazard rates $m$ and $M$ that enter the utility calculation, but rather the individual's \emph{beliefs} about these rates. Particularly, in the case of human extinction risk $M$, which, by definition, can only materialize once in human history and has never materialized yet, these beliefs are highly speculative, subjective, and uncertain.

\begin{table}\label{tab:results}
\centering
\caption{Summary of analytical results}
\begin{tabular}{lll}
\hline
Case & Discount factor & Discount factor if $n=0$ \\
\hline
Individual utility & $(1-M)(1-m)$ & $(1-M)(1-m)$ \\
Utility of a dynasty & $(1-M)(1-m)(1+b)$ & $1-M$ \\
Utility of a dynasty ($\theta\in[0,1]$) & $(1-M)(1-m)^\theta(1+b)^\theta$ & $1-M$ \\
Utility of a genetic lineage & $(1-M)(1-m)(1+b)^\alpha$ & $(1-M)(1-m)^{1-\alpha}$ \\
Social welfare & $(1-M)(1-m)(1+b)$ & $1-M$ \\
\hline
\end{tabular}

\end{table}

With these results in hand, we can address the questions raised in the Introduction:
\begin{itemize}
\item \emph{When people update their beliefs on the underlying extinction risk upwards, do they become more impatient?}
Our results are highly suggestive of that. The perceived extinction risk is a major source of impatience, and one that cannot be hedged against through reproduction. Therefore it affects the discount rate equally irrespective of the extent of individuals' intergenerational altruism.

\item \emph{If so, will this increase in impatience be smaller or larger than the increase in impatience observed among individuals who update their beliefs about their own expected lifespan, for example when they are diagnosed with a terminal illness?} 
According to our model, this increase in impatience will be either equal or greater than the corresponding increase in impatience observed among individuals who update their beliefs about their own mortality. Specifically, with a positive extent of intergenerational altruism, and assuming that the population growth rate $n$ is predetermined, the response to an increase in perceived extinction risk should be relatively greater because extinction risk---as opposed to individual mortality risk---cannot be hedged against through reproduction.

\item \emph{If rising extinction risks lead to higher discounting, what does this imply for investment into the prevention of extinction in the first place?} 
In the case of a rise in the discount rate, caused by the acknowledgment of higher extinction risk which cannot be hedged against, humanity would become more present-focused. Then it would be even harder to convince us to invest in projects that could mitigate extinction risks themselves. This is a destabilizing force because once humanity deviates from a sustainable path of long-run development towards a path with increased extinction risk, the endogenous rise in the discount rate caused by this deviation would make it even more difficult to get back to the sustainable path. The only exception to this self-defeating logic would be if humanity believed that a given intervention would successfully lower the underlying extinction risk, which would allow us to approve the intervention and simultaneously adjust our beliefs on extinction risk back down.

\end{itemize}


We conclude that if people were to update their beliefs on the risk of human extinction upwards---whether in response to the recent developments in AI, advancing climate change, or otherwise---they will likely also become more impatient. This is in itself destabilizing because with greater patience, humanity would be more willing to invest in extinction risk reduction. 

The current state of affairs is that extinction risk seems to be widely neglected in political decision-making, as if (counterfactually) $M \approx 0$, and as if no investment in extinction risk reduction were needed whatsoever. This may be due to a deeply rooted human psychological inclination to deny the existence of risk factors that have not visibly materialized in the past; unfortunately such an instinct is self-defeating because it does not allow for belief updating after the extinction event. But as our results indicate, acknowledging the presence of extinction risk need not be helpful and may be even counterproductive, because it could bring about an endogenous increase in our impatience---another force that prevents societies and policymakers from acting appropriately. We speculate that this mechanism could only be counteracted if humanity were successfully convinced that a certain proposed intervention (e.g., pausing the development of dangerous AI capabilities, investing in certain lines of safety research) would successfully lower the extinction risk in the years to come.

\section{Conclusion}
\label{sec:conclusions}

We have shown that extinction risk is a fundamental source of impatience that cannot be hedged away through reproduction in the context of intergenerational altruism. Unlike individual mortality risk in many specifications, it directly increases discounting, thereby reducing the attractiveness of long-term investments. This gives rise to a paradox: the greater the danger of extinction is, the weaker are the incentives to prevent it.

This mechanism offers an explanation for the persistent underinvestment in extinction risk mitigation ranging from climate change to pandemic prevention and the potential risks associated with transformative AI. Rising extinction risk endogenously induces short-termism, which, in turn, makes societies less likely to address the very risks that threaten them. Consequently, our results highlight the need for institutional arrangements that embed more farsightedness into collective decision-making. Internationally binding carbon pricing schemes, international coordination on AI safety, and binding and enforced treaties on limiting biological, chemical, and nuclear weapons could be helpful in counteracting the dynamics that we identify.

\bibliographystyle{apalike}
\bibliography{bibliography}

\begin{thebibliography}{}

\bibitem[Blanchard, 1985]{Blanchard1985}
Blanchard, O.~J. (1985).
\newblock {Debt, Deficits, and Finite Horizons}.
\newblock {\em Journal of Political Economy}, 93:223--247.

\bibitem[Bostrom, 2002]{Bostrom2002}
Bostrom, N. (2002).
\newblock {Existential Risks: Analyzing Human Extinction Scenarios and Related
  Hazards}.
\newblock {\em Journal of Evolution and Technology}, 9.

\bibitem[Bostrom, 2014]{Bostrom2014}
Bostrom, N. (2014).
\newblock {\em {Superintelligence: Paths, Dangers, Strategies}}.
\newblock Oxford University Press.

\bibitem[Chichilnisky, 2000]{Chichilnisky2000}
Chichilnisky, G. (2000).
\newblock {An Axiomatic Approach to Choice Under Uncertainty with Catastrophic
  Risks}.
\newblock {\em Resource and Energy Economics}, 22:221--231.

\bibitem[Chichilnisky et~al., 2020]{ChichilniskyEtal2020}
Chichilnisky, G., Hammond, P.~J., and Stern, N. (2020).
\newblock {Fundamental Utilitarianism and Intergenerational Equity With
  Extinction Discounting}.
\newblock {\em Social Choice and Welfare}, 54:397--427.

\bibitem[Cohen et~al., 2020]{CohenEtal2020}
Cohen, J., Ericson, K.~M., Laibson, D., and White, J.~M. (2020).
\newblock {Measuring Time Preferences}.
\newblock {\em Journal of Economic Literature}, 58:299--347.

\bibitem[Cooley and Prescott, 1995]{CooleyPrescott1995}
Cooley, T. and Prescott, E.~C. (1995).
\newblock {Economic Growth and Business Cycles}.
\newblock In Cooley, T., editor, {\em {Frontiers of Business Cycle Research}},
  pages 1--38. Princeton University Press, Princeton, NJ.

\bibitem[Dawkins, 1976]{Dawkins1976}
Dawkins, R. (1976).
\newblock {\em {The Selfish Gene}}.
\newblock Best Books.

\bibitem[Feng and Ke, 2018]{Feng2018}
Feng, T. and Ke, S. (2018).
\newblock {Social Discounting and Intergenerational Pareto}.
\newblock {\em Econometrica}, 86(5):1537--1567.

\bibitem[Fisher, 1930]{Fisher1930}
Fisher, I. (1930).
\newblock {\em {The Theory of Interest}}.
\newblock New York: Macmillan.

\bibitem[Frederick et~al., 2002]{FrederickEtal2002}
Frederick, S., Loewenstein, G., and O’Donoghue, T. (2002).
\newblock {Time Discounting and Time Preference: A Critical Review}.
\newblock {\em Journal of Economic Literature}, 40:351--401.

\bibitem[Gomme and Rupert, 2007]{GommeRupert2007}
Gomme, P. and Rupert, P. (2007).
\newblock {Theory, Measurement and Calibration of Macroeconomic Models}.
\newblock {\em Journal of Monetary Economics}, 54:460--497.

\bibitem[Growiec and Prettner, 2025]{GrowiecPrettner2025}
Growiec, J. and Prettner, K. (2025).
\newblock {The Economics of p(doom): Scenarios of Existential Risk and Economic
  Growth in the Age of Transformative AI}.
\newblock arxiv:2503.07341, SGH Warsaw School of Economics.

\bibitem[Hayden, 2016]{Hayden2016}
Hayden, B.~Y. (2016).
\newblock {Time Discounting and Time Preference in Animals: A Critical Review}.
\newblock {\em Psychonomic Bulletin and Review}, 23:39--53.

\bibitem[Imai et~al., 2021]{ImaiEtal2021}
Imai, T., Rutter, T.~A., and Camerer, C.~F. (2021).
\newblock {Meta-Analysis of Present-Bias Estimation Using Convex Time Budgets}.
\newblock {\em Economic Journal}, 131:1788--1814.

\bibitem[Jones, 2023]{Jones2023}
Jones, C.~I. (2023).
\newblock {The AI Dilemma: Growth versus Existential Risk}.
\newblock {NBER Working Paper 31837}, National Bureau of Economic Research.

\bibitem[Martin and Pindyck, 2015]{MartinPindyck2015}
Martin, I. and Pindyck, R.~S. (2015).
\newblock {Averting Catastrophes: The Strange Economics of Scylla and
  Charybdis}.
\newblock {\em American Economic Review}, 105:2947--2985.

\bibitem[Matousek et~al., 2022]{MatousekEtal2022}
Matousek, J., Havranek, T., and Irsova, Z. (2022).
\newblock {Individual Discount Rates: A Meta‑Analysis of Experimental
  Evidence}.
\newblock {\em Experimental Economics}, 25:318--358.

\bibitem[Mischel et~al., 1989]{MischelEtal1989}
Mischel, W., Shoda, Y., and Rodriguez, M. (1989).
\newblock {Delay of Gratification in Children}.
\newblock {\em Science}, 244:933--938.

\bibitem[Ord, 2020]{Ord2020}
Ord, T. (2020).
\newblock {\em {The Precipice: Existential Risk and the Future of Humanity}}.
\newblock Hachette.

\bibitem[Parfit, 1986]{Parfit1986}
Parfit, D. (1986).
\newblock {\em {Reasons and Persons}}.
\newblock Oxford University Press.

\bibitem[Robson, 2001]{Robson2001}
Robson, A.~J. (2001).
\newblock {The Biological Basis of Economic Behavior}.
\newblock {\em Journal of Economic Literature}, 39:11--33.

\bibitem[Robson and Samuelson, 2009]{RobsonSamuelson2009}
Robson, A.~J. and Samuelson, L. (2009).
\newblock {The Evolution of Time Preference with Aggregate Uncertainty}.
\newblock {\em American Economic Review}, 99:1925--1953.

\bibitem[Robson and Samuelson, 2010]{RobsonSamuelson2010}
Robson, A.~J. and Samuelson, L. (2010).
\newblock {The Evolutionary Foundations of Preferences}.
\newblock In Benhabib, J., Bisin, A., and Jackson, M., editors, {\em {The
  Social Economics Handbook}}. Elsevier.

\bibitem[van~der Wal et~al., 2013]{VanDerWalEtal2013}
van~der Wal, A.~J., Schade, H.~M., Krabbendam, L., and van Vugt, M. (2013).
\newblock {Do Natural Landscapes Reduce Future Discounting in Humans?}
\newblock {\em Proceedings of the Royal Society B}, 280:20132295.

\bibitem[Vanderveldt et~al., 2016]{VanderveldtEtal2016}
Vanderveldt, A., Oliveira, L., and Green, L. (2016).
\newblock {Delay Discounting: Pigeon, Rat, Human---Does It Matter?}
\newblock {\em Journal of Experimental Psychology: Animal Learning and
  Cognition}, 42:141--162.

\bibitem[Weitzman, 2012]{Weitzman2012}
Weitzman, M.~L. (2012).
\newblock {GHG Targets as Insurance Against Catastrophic Climate Damages}.
\newblock {\em Journal of Public Economic Theory}, 14(2):221--244.

\bibitem[Yudkowsky and Soares, 2025]{YudkowskySoares2025}
Yudkowsky, E. and Soares, N. (2025).
\newblock {\em {If Anyone Builds It, Everyone Dies}}.
\newblock Little, Brown and Company.

\end{thebibliography}

\end{document}